\documentclass[12pt,preprint]{aastex}

\def\24m{24 $\mu$m}
\def\lir{$L_{IR}$ }

\def\sm{$M_{*}$ }
\def\kpc{$h^{-1}$kpc }
\def\kms{${\rm km~s^{-1}}$}
\def\speir{$L_{IR}/M_{*}$ }
\def\dis{$\vartriangle$$r_{p}$ }
\def\vel{$\vartriangle$$v$ }
\def\solar{$L_{\odot}$}

\slugcomment{Accepted for publication in the ApJL AEGIS Special Issue}

\shorttitle{Star Formation in Close Galaxy Pairs and Merging Galaxies}
\shortauthors{Lin et al.}

\begin{document}

\title{AEGIS: Enhancement of Dust-enshrouded Star Formation in Close Galaxy Pairs and Merging Galaxies up to $z \sim 1$ \altaffilmark{1}}

\author{Lihwai Lin \altaffilmark{2,3}, David C. Koo \altaffilmark{3}, Benjamin J. Weiner \altaffilmark{4}, Tzihong
Chiueh\altaffilmark{2}, Alison L. Coil \altaffilmark{5,6},
Jennifer Lotz \altaffilmark{7,8}, Christopher J. Conselice
\altaffilmark{9}, S. P. Willner \altaffilmark{10}, H. A. Smith
\altaffilmark{10}, Puragra Guhathakurta \altaffilmark{3}, J.-S.
Huang \altaffilmark{10}, Emeric Le Floc'h \altaffilmark{5}, Kai G.
Noeske \altaffilmark{3}, Christopher N. A. Willmer
\altaffilmark{5}, Michael C. Cooper \altaffilmark{11}, and Andrew
C. Phillips \altaffilmark{3}}

\altaffiltext{1}{ Some of the data presented herein were obtained at the W. M. Keck Observatory, which is operated
as a scientific partnership among the California Institute of Technology, the University of California and the
National Aeronautics and Space Administration. The Observatory was made possible by the generous financial support
of the W.M. Keck Foundation.} \altaffiltext{2}{Department of Physics, National Taiwan University, Taipei 106,
Taiwan; Email: d90222005@ntu.edu.tw} \altaffiltext{3}{UCO/Lick Observatory, Department of Astronomy and
Astrophysics, University of California, Santa Cruz, CA 95064} \altaffiltext{4}{Department of Astronomy, University
of Maryland, College Park, MD  20742} \altaffiltext{5}{Steward Observatory, University of Arizona, Tucson, AZ 85721}
\altaffiltext{6}{Hubble Fellow} \altaffiltext{7}{National Optical Astronomy Observatory, Tucson, AZ 85719}
\altaffiltext{8}{Leo Goldberg Fellow} \altaffiltext{9}{School of Physics and Astronomy, University of Nottingham,
Nottingham NG7 2RD, UK} \altaffiltext{10}{Harvard-Smithsonian Center for Astrophysics, Cambridge, MA 02138}
\altaffiltext{11}{Department of Astronomy, University of California, Berkeley, CA 94720}

\begin{abstract}
\hspace{3mm} Using data from the DEEP2 Galaxy Redshift Survey and \textit{HST}/ACS imaging in the Extended Groth
Strip, we select nearly 100 interacting galaxy systems including kinematic close pairs and morphologically
identified merging galaxies. $Spitzer$ MIPS \24m fluxes of these systems reflect the current dusty star formation
activity, and at a fixed stellar mass (\sm) the median infrared luminosity (\lir) among merging galaxies and close
pairs of blue galaxies is twice (1.9$\pm$0.4) that of control pairs drawn from isolated blue galaxies. Enhancement
declines with galaxy separation, being strongest in close pairs and mergers and weaker in wide pairs compared to the
control sample. At $\overline{z} \sim 0.9$, $7.1\%\pm4.3\%$ of massive interacting galaxies (\sm $>$ $2\times10^{10}
M_{\odot}$) are found to be ULIRGs, compared to $2.6\%\pm0.7\%$ in the control sample. The large spread of \speir
among interacting galaxies suggests that this enhancement may depend on the merger stage as well as other as yet
unidentified factors (e.g., galaxy structure, mass ratio, orbital characteristics, presence of AGN or bar). The
contribution of interacting systems to the total IR luminosity density is moderate ($\la 36\%$).

\end{abstract}


\section{INTRODUCTION}

Galaxy-galaxy interaction has long been regarded as a key process in galaxy evolution, especially as a mechanism for
enhancing star formation during mergers \citep{lar78, bar00}. Hydrodynamic $N$-body simulations show that active
star formation can be triggered by gaseous inflows resulting from mergers of gas-rich galaxies \citep{mih96,
bar04,cox04}. Interaction-triggered star formation is also thought to be responsible for luminous infrared sources.
In the local universe, luminous infrared galaxies (LIRGs) and ultra luminous infrared galaxies (ULIRGs) are
primarily merging systems \citep{san88,bor99}. Nevertheless, by studying various star formation indicators of
interacting galaxies and normal galaxies, \citet{ber03} concluded that galaxy interactions in general are
inefficient triggers of starbursts; interactions are a necessary but not sufficient condition to trigger violent
starbursts.

The importance of galaxy interactions in the volume-averaged
galaxy star formation rate (SFR) remains an open question. Recent
studies of mid-IR (MIR) sources at a median redshift of $z\sim$0.7
suggest that the IR density at that epoch is dominated by
morphologically normal galaxies instead of strongly interacting
galaxies \citep{bel05,mel05}. Two main factors may contribute to
this result. First, only a small fraction of the galaxy population
may be undergoing a major merger at any given time
\citep{car00,bun04,lin04}. Second, the overall SFR in normal
galaxies at $z\sim$0.7 may be enhanced relative to the local
population \citep{bel05}, perhaps as the result of internal
processes such as a higher gas fraction leading to a higher SFR,
such that galaxy interactions may have less dramatic effects on
triggering star formation at that epoch and/or may be harder to
identify. It is the aim of this Letter to examine this second
hypothesis.

This Letter presents an analysis of the IR properties of close kinematic galaxy pairs, morphologically selected
merging galaxies, and a control sample of randomly selected pairs of isolated galaxies, in the range $0.1<z<1.1$. In
$\S$3 we show our analysis of the IR luminosity (\lir) versus stellar mass (\sm) for the interacting galaxies and
control samples, and the relation between the IR luminosity-to-mass ratio (\speir) and the projected separation of
the galaxy pairs. Discussion and conclusions are given in $\S$4. Throughout this Letter we adopt the following
cosmology: H$_0$ = 100$h$~\kms Mpc$^{-1}$, $\Omega_m = 0.3$, and $\Omega_{\Lambda} = 0.7$; $h$ = 0.7 is adopted when
calculating the rest-frame magnitude. Magnitudes are given in the Vega system.

\section{DATA, SAMPLE SELECTIONS, AND METHODS}
\subsection{Data}
The sample used in this Letter consists of $\sim$ 4000 optical selected galaxies with secure redshifts from the
DEEP2 Galaxy Redshift Survey and MIPS imaging in the Extended Groth Strip (EGS) region. The EGS has extensive
multi-wavelength observations from both the ground and space, coordinated by the All-Wavelength Extended Groth
International Survey (AEGIS) team \citep{dav06}. The spectral resolution of the DEEP2 survey is R $\sim$ 5000,
corresponding to redshift errors of $\sim$30 \kms, allowing us to define kinematic pairs (see $\S$ 2.2). The average
sampling rate of the slit placed on galaxies is $\sim 60\%$, and the average redshift success is $\sim 73\%$
\citep{wil06}. Details of the survey and $K$-correction procedure are described in a series of DEEP2 papers
\citep{dav03, coi04,wil06}. The MIPS (Multiband Imaging Photometer for Spitzer; Rieke et al. 2004) \24m observations
used here were carried out on 2004 June 19 and 20; \citet{dav06} provide further details.

\subsection{Selection of Kinematic Pairs, Merging Galaxies, and Control Samples}
Three classes of galaxy samples are investigated in this study: kinematic pairs, merging galaxies, and a control
sample of randomly selected pairs of isolated galaxies, in three redshift bins in the range $0.1<z<1.1$. To select
\textbf{kinematic galaxy pairs}, we use the following criteria: (1) For each galaxy in the spectroscopic redshift
sample, we first search for any kinematic companions with a relative line-of-sight velocity \vel $\leq$ 500 \kms and
a projected physical separation (onto the plane of the sky) \dis $<$300 \kpc. (2) Among these companions, close
pairs are identified such that they satisfy \dis $\leq$ 50 \kpc \citep{pat02,lin04}. We find a total of 56 close
pairs. (3) For comparison, wide pairs are identified as kinematic companions with 50 \kpc $\leq$ \dis $\leq$ 300
\kpc, and we retain only wide pairs that are closest companions to each other (such that there is only one wide pair
companion per galaxy). The spectroscopic redshift sample is not entirely complete, however; it is possible that for
a given galaxy, there exists a closer companion that is missing in the redshift sample. Therefore, \dis for each
paired system is an upper limit on the actual distance to the closest companion. To minimize this effect we also
search for photometric companions of each galaxy, and we keep only those wide pairs that do not have a photometric
companion (with observed $\vartriangle$$m_{R} <$ 1 mag) within \dis $\leq$ 30 \kpc, for a sample of 126 wide pairs.
(4) Absolute magnitude limits were applied to all pair samples within a given redshift bin (see Table 1). (5) To
study the effect of interactions between two gas-rich galaxies, we further applied a color cut using rest-frame $U -
B$ = 0.25 as the division between blue and red galaxies and only include blue galaxies in our samples. It is
possible that we have missed some red, dusty, late-type galaxies due to our color selection. However, the number
density of that population is $\sim$8\% compared to the blue galaxy population \citep{wei05} and is therefore
negligible for our analysis.

We also define a set of morphologically identified \textbf{merging galaxies} using a subsample of blue galaxies in
the EGS that have deep $HST$ images taken with Advanced Camera for Surveys (ACS) as part of GO program 10134 (PI: M.
Davis). These morphologically identified merging galaxies are preferentially in strongly interacting systems and are
in a different merger stage than the kinematic close pairs. We identify merging galaxies using three non-parametric
parameters: the Gini coefficient ($G$), the second-order moment of the brightest 20\% of a galaxy's pixels
($M_{20}$), and the asymmetry measurement ($A$), all of which have been shown to identify merger candidates
\citep{con03a,con03b,lot04}. We first select galaxies with morphological parameters of $G
> -0.115*M_{20} + 0.384$ or $A>0.25$ (see Lotz et al. 2007 for
discussion on $G$ and $M_{20}$ in detail), and then we perform
by-eye examinations of each object to keep only those candidates
that show apparent interaction signatures (eg., tidal tails,
distorted morphology, and double nuclei). This results in a sample
of 56 merging systems.

For a fair comparison with isolated galaxies we also construct a sample of 1800 \textbf{control pairs}, each of
which consists of two galaxies randomly selected from galaxies that are isolated, with no spectroscopic companion
within 100 \kpc or photometric companion (with observed $\vartriangle$$m_{R}$ $<$ 1 mag) within 30 \kpc. Isolated
galaxies have the same magnitude and color cut as adopted for interacting galaxies.

\subsection{Stellar Mass and Total IR Luminosity}
Stellar masses are derived from rest-frame $(B-V)$ colors and absolute $M_B$ magnitude as described by the models of
\citet{bel01} and \citet{bel05}, who find a scatter of $\sim$0.3 dex in the resulting stellar mass estimates. These
measurements were further refined by comparison with the stellar masses derived by \citet{bun05} using detailed fits
to the spectral energy distribution (SED).  Empirically we find that the difference between the masses estimated
using the rest-frame colors and the SED fits is improved by making small corrections for the redshift, rest-frame
$(U-B)$ and $(B-V)$ colors:
\begin{eqnarray}
log_{10}\frac{M_{*}}{M_{\odot}}&=&-0.4(M_{B}-5.48)+1.737(B-V)+0.098(U-B)\nonumber\\&&-0.130(U-B)^{2}-0.268z-1.003.
\end{eqnarray}
These terms have the effect of correcting the $z=0$ measurements of \citet{bel01} to the galaxy redshift as well as
accounting for evolution in color. Masses estimated from Equation 1 agree with those from full SED fits (when
available; Bundy et al. 2006) to an rms accuracy of 0.25 dex.

Local studies have shown that the rest-frame MIR luminosity is tightly correlated with the total IR luminosity over
a wide range of galaxy types \citep{rou01, pap02}. This relation appears to hold up reasonably well to $z \sim 1$
\citep{elb02, app04} and hence allows us to estimate the IR luminosity from the observed flux of \24m corresponding
to the rest-frame 11-21 $\mu$m flux over the redshift range $0.1<z<1.1$. Following the same procedure adopted by
\citet{lef05}, we convert the observed \24m flux into the total IR luminosity by fitting for each source using SEDs
from \citet{cha01}. \footnote{Various template libraries lead to a scatter in this conversion of a factor of 2-3
\citep{lef05}.} Because of the large PSF (FWHM 6") of the \24m data, we work on the total IR luminosity in pairs
rather than the IR luminosity of each individual galaxy. This also allows us to make comparisons to local ULIRGs,
which are found to be highly distorted merging systems \citep{san88,bor99}, and to merger simulations in the
literature \citep{bar04,jon06}, which yield the total star formation rate of the merger system. For each pair, we
then calculate the mean IR luminosity-to-mass ratio, \speir, by summing the IR luminosity of both galaxies and
dividing by the total stellar mass of both components.

\section{RESULTS}

Figure 1 shows the relation between \lir and \sm for three of the galaxy samples.  A relatively tight correlation is
seen for the control pairs, and the close pairs and mergers are within the bounds of this relation but tend to
occupy the upper region of \lir for a given \sm. In the highest redshift bin, the fraction of merging galaxies and
close pairs more massive than $2\times10^{10} M_{\odot}$ that are ULIRGs is $7.1\%\pm4.3$\%, compared to
$2.6\%\pm0.7$\% for the control sample. To address quantitatively the difference of \lir among those samples, we
compute the median \speir for each sample. Using the median value of \speir rather than performing the K-S test
avoids the problem of handling the upper limits. When calculating the median \speir, stellar mass cuts are further
applied in each redshift bin (see Table 1). Over the entire redshift range, the median \speir of pairs and mergers
is twice ($1.9\pm0.4$) that of the control sample. To assess if our results are sensitive to the method we have
adopted to estimate stellar masses, we also apply the same analysis on a subsample of the data with stellar mass
measurements derived from the SED-fitting procedure of \cite{bun05}. The enhancement in \speir increases to 3.1
$\pm$1.3 in this case, consistent with our initial result within the $1 \sigma$ uncertainties.

An alternative way to investigate the effect of galaxy interactions on the star formation rate is to study the
dependence of \speir on the pair separation, as shown in Figure 2. For comparison, distributions of \speir in
control pairs are also shown along the right-hand axes. The merging galaxies and close pairs possess higher median
\speir than wide pairs and control pairs, as shown in Table 1, but also have a wider spread in \speir. The declining
envelope of \speir as a function of \dis is similar to the behavior of the local pair results reported by
\citet{bar00}, who use various sets of emission lines as the star formation tracer. The median \speir of wide pairs,
on the other hand, becomes close to that of the control pairs, indicating that the effect of galaxy interaction on
star formation activity is limited to several tens of \kpc.

\section{DISCUSSION AND CONCLUSION }
Using data from the DEEP2 Galaxy Redshift Survey with MIPS \24m imaging in the EGS, we find that the combined IR
luminosity at a given stellar mass of blue merging galaxies and kinematic close pairs is greater by a factor of $1.9
\pm 0.4$ than that of control pairs randomly drawn from blue isolated galaxies. This enhancement is consistent with
low-redshift studies of the SFR in galaxy pairs \citep{lam03,nik04}. In addition, we also observe a declining
envelope of the \speir with increased projected separation of kinematic galaxy pairs.
Based on the assumption that the IR emission is tightly associated with the SFR, our results qualitatively support
the picture of tidally triggered starbursts as predicted in hydrodynamic simulations, although the effect is
apparently not large. The frequency of ULIRGs in massive interacting systems is only $7.1\%\pm4.3\%$ in the highest
redshift bin ($0.75 < z < 1.1$), and no ULIRGs are found below a combined stellar mass for the pair of
$4\times10^{10} M_{\odot}$. The mass ratio between two gas-rich galaxies may be a key element in generating ULIRG
luminosities \citep{das06}, along with the dependence on the combined stellar mass of the interacting system. The
wide spread of specific star formation rates in interacting galaxies found here (as inferred from the broad
distribution of \speir), and also seen in \citet{bar00}, indicates that some other mechanism(s) may determine the
strength of the induced star formation as well -- e.g., the galaxy structure, the mass/luminosity ratio, the orbital
configuration, and the existence of AGN and/or bar. For example, studies by \citet{lam03} and \citet{woo06} suggest
that pairs with comparable luminosity do show stronger star formation activity than pairs with a larger luminosity
contrast.  The scatter in \speir that we find could also be due to the fact that these systems are in different
stages of merging: some kinematic pairs are likely on their first approach while others are being seen after their
first passage. Another possibility is that some of the close pairs selected by our \dis and \vel criteria are not
physical close pairs in real space (as opposed to redshift space; Perez et al. 2006). Our results nevertheless
provide a constraint on the amount of induced star formation activity due to galaxy interactions at $z \sim
0.1-1.1$.

Finally, we note that tidally triggered star formation contributes moderately to the high IR luminosity density at
intermediate redshifts of $0.4<z<1.1$. \citet{lin04} estimated that the pair fraction (\dis$\leq$50 \kpc and
\vel$\leq$500 \kms) of $L^{*}$ galaxies over this redshift range is less than 15\% and that the redshift evolution
in the pair fraction is also much lower than the rapid decline of IR luminosity density seen with decreasing
redshift. Additionally, the fraction of morphologically identified merging galaxies remains roughly constant at 7\%
up to $z \sim 1$ \citep{lot06}. These results, when combined with the moderate increase of star formation activity
seen in interacting systems shown here, indicate that the contribution of galaxy interactions to the total IR
density at intermediate redshift is $\la 36\%$. This is consistent with the conclusions from the studies of IR
populations of various morphology types by \citet{bel05} and \citet{mel05}.

\acknowledgments This Letter was prepared as part of the All-Wavelength Extended Groth International Survey (AEGIS)
collaboration. We thank S. M. Faber and P. Jonsson for useful discussions and the anonymous referee for valuable
comments; L. Lin acknowledges support from Taiwan COSPA project, and from Taiwan NSC grant NSC94-2112-M-002-026 and
NSC93-2917-I-002-008. C. J. Conselice acknowledge an NSF Astronomy and Astrophysics Fellowship and PPARC for
support. A. L. Coil is supported by NASA through Hubble Fellowship grant HF-01182.01-A. The DEEP2 project was
supported by NSF grants AST00-71198, AST05-07428 and AST05-07483. This work is based in part on observations made
with the \textit{Spitzer Space Telescope}, which is operated by the Jet Propulsion Laboratory, California Institute
of Technology, under a contract with NASA. Support for this work was provided by NASA through contract 1255094
issued by JPL/Caltech. We close with thanks to the Hawaiian people for use of their sacred mountain.

\clearpage

\begin{deluxetable}{ccrrrrrrr}
\tabletypesize{\scriptsize} \tablecaption{Properties of the pair sample \label{tbl-1}} \tablewidth{0pt} \tablehead{
\colhead{$z$} & \colhead{Sample} & \colhead{$M_{B}$(individual)} &\colhead{$M_{*}$(combined)} & \colhead{$No.$} &
\colhead{$d_{24}$} & \colhead{median \speir}} \startdata
0.1 $< z <$0.4 &MG &-21.7 $\leq$ $M_{B}$ $\leq$ -18.7 &9.7$\leq$log$_{10}M_{*}$$\leq$10.7 &3(9) &\nodata &\nodata\\
\nodata &close pair &-21 $\leq$ $M_{B}$ $\leq$ -18 &9.7$\leq$log$_{10}M_{*}$$\leq$10.7 &10(10) &68$\pm$10(\%) &2.2$\pm$0.7\\
\nodata &wide pair &-21 $\leq$ $M_{B}$ $\leq$ -18 &9.7$\leq$log$_{10}M_{*}$$\leq$10.7 &17(20) &89$\pm$6(\%) &1.3$\pm$0.1\\
\nodata &control pair &-21 $\leq$ $M_{B}$ $\leq$ -18 &9.7$\leq$log$_{10}M_{*}$$\leq$10.7 &480(600) &74$\pm$11(\%) &1.0$\pm$0.3\\
\tableline
0.4 $< z <$0.75 &MG &-22.7 $\leq$ $M_{B}$ $\leq$ -19.7 &10$\leq$log$_{10}M_{*}$$\leq$11 &13(18)&62$\pm$13(\%) &3.1$\pm$0.8\\
\nodata &close pair &-22 $\leq$ $M_{B}$ $\leq$ -19 &10$\leq$log$_{10}M_{*}$$\leq$11 &21(24) &71$\pm$7(\%) &2.7$\pm$0.4\\
\nodata &wide pair &-22 $\leq$ $M_{B}$ $\leq$ -19 &10$\leq$log$_{10}M_{*}$$\leq$11 &44(54) &64$\pm$5(\%) &1.7$\pm$0.3\\
\nodata &control pair &-22 $\leq$ $M_{B}$ $\leq$ -19 &10$\leq$log$_{10}M_{*}$$\leq$11 &502(600) &60$\pm$7(\%) &1.5$\pm$0.4\\
\tableline
0.75 $< z <$1.1 &MG &-23.7 $\leq$ $M_{B}$ $\leq$ -20.7 &10.3$\leq$log$_{10}M_{*}$$\leq$11.3 &20(29) &73$\pm$10(\%) &3.3$\pm$2.7\\
\nodata &close pair &-23 $\leq$ $M_{B}$ $\leq$ -20 &10.3$\leq$log$_{10}M_{*}$$\leq$11.3 &22(22) &76$\pm$7(\%) &3.8$\pm$0.8\\
\nodata &wide pair &-23 $\leq$ $M_{B}$ $\leq$ -20 &10.3$\leq$log$_{10}M_{*}$$\leq$11.3 &45(52) &62$\pm$5(\%) &1.5$\pm$0.2\\
\nodata &control pair &-23 $\leq$ $M_{B}$ $\leq$ -20 &10.3$\leq$log$_{10}M_{*}$$\leq$11.3 &547(600) &64$\pm$7(\%) &1.9$\pm$0.5\\
\enddata
\tablecomments{ $M_{B}$ is the rest-frame $B$-band magnitude of each galaxy in pairs. The luminosity range of
merging galaxies (MGs) is chosen to be twice as bright as that of individual galaxy in pairs because merging
galaxies are assumed to be comprised of two galaxies.
\sm is the stellar mass cut when computing the median \speir and \24m detection rate $d_{24}$. In the fifth column,
the values outside (inside) the parentheses are the numbers of paired systems with (without) a stellar mass cut.
Error bars quoted here are computed by bootstrapping.}
\end{deluxetable}

\clearpage

\begin{figure}
\epsscale{.7} \plotone{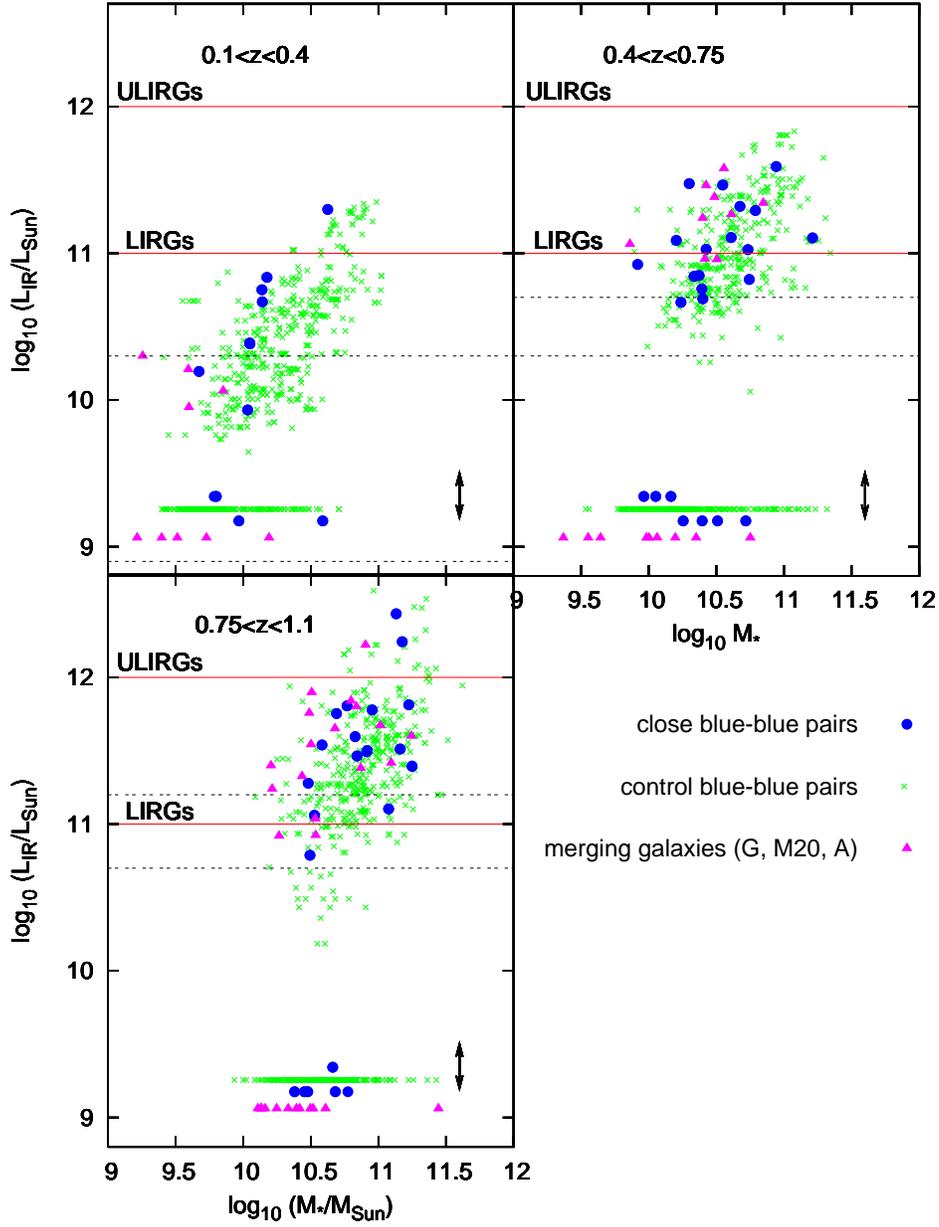} \caption{Total IR luminosity (\lir) vs. stellar mass (\sm) for kinematic close pairs
(blue circles), control pairs (green crosses), and morphologically identified merging galaxies (magenta triangles).
Systems without a \24m detection are assigned low \lir values $<3\times10^{9}$ \solar. Horizontal red solid lines
represent the IR thresholds for ULIRGs ($L_{IR}$ $\geq$ $10^{12}$ \solar) and LIRGs ($L_{IR}$ $\geq$ $10^{11}$
\solar). Black dotted lines correspond to 70 $\mu$Jy (5$\sigma$) in \24m at lower and upper redshift limits. Arrows
in the lower right corner of each panel show the scatter (0.3 dex) in the conversion of \24m flux into total \lir
between different template libraries. It can been seen that the close pairs and merging galaxies possess higher
$L_{IR}$ than control pairs for a given \sm. \label{fig1}}
\end{figure}

\clearpage

\begin{figure}
\epsscale{.7} \plotone{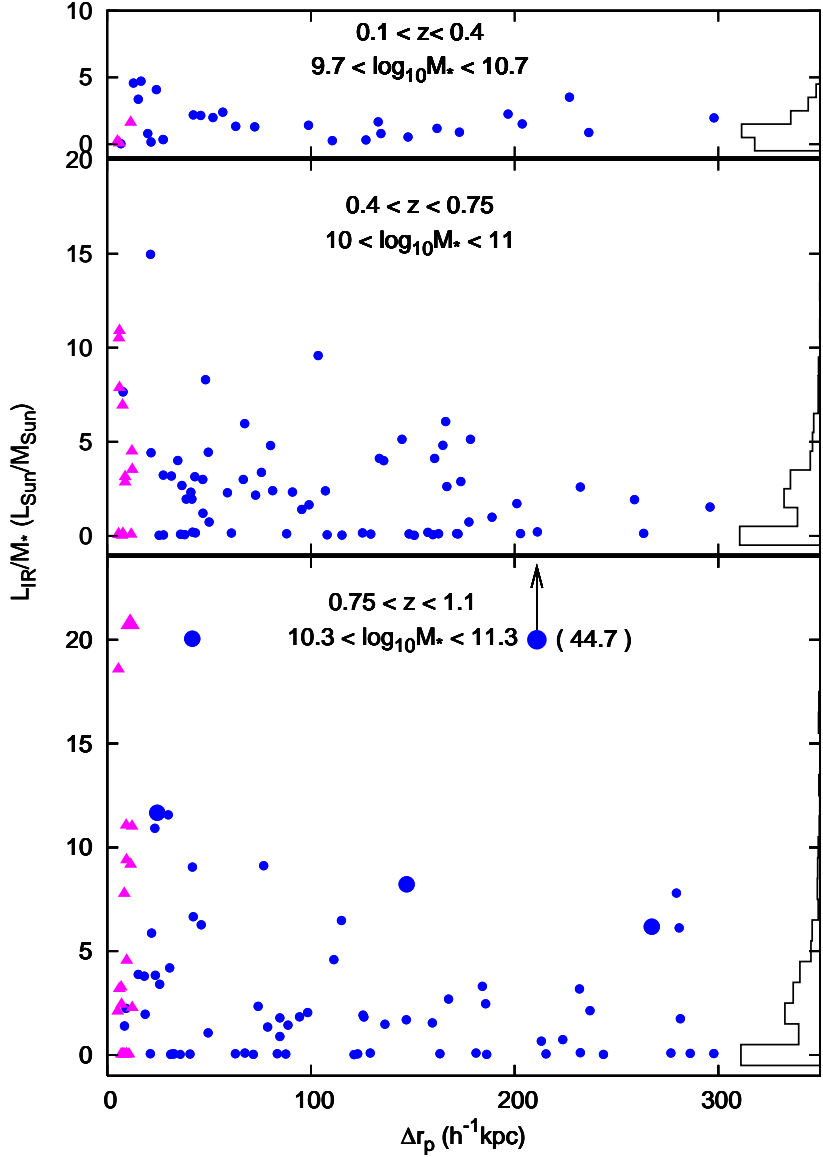} \caption{\speir as a function of projected separation for kinematic galaxy pairs
(blue circles) and merging galaxies (magenta triangles) in three redshift bins. Merging galaxies are assigned to
\dis $\sim5$ \kpc, which is the minimum separation of kinematic close pairs. Galaxies with $L_{IR}$ $>$ $10^{12}$
$L_{\odot}$ (ULIRGs) are shown as larger symbols. Distributions of \speir in control pairs are also shown along the
right-hand axes for comparison. Data points at \speir = 0 refer to those sources with no \24m detection. The number
within the parentheses in the bottom panel denotes the \speir of the closest data point. Close pairs (\dis $<$ 50
\kpc) and merging galaxies are found to have a higher median \speir and a wider spread of \speir than those of wide
pairs (\dis $>$ 50 \kpc) or control pairs (see Table 1). \label{fig2}}
\end{figure}

\end{document}